\documentclass[12pt]{article}
\usepackage{graphicx}
\usepackage{amssymb}
\usepackage{amsmath}
\newcommand{\vev}[1]{{\langle #1 \rangle}}

\newcommand{\eV}{\mbox{~eV}}
\newcommand{\MeV}{\mbox{~MeV}}

\newcommand{\ie}{{\it i.e.}}
\newcommand{\eg}{{\it e.g.}}
\newcommand{\eqn}[1]{&\hspace{-0.6em}#1\hspace{-0.6em}&}
\begin{document}
\baselineskip 0.6cm
%
\begin{titlepage}
\begin{center}

\begin{flushright}
\end{flushright}

\vskip 2cm

{\Large \bf Direct Search for Right-handed Neutrinos \\
and Neutrinoless Double Beta Decay}

\vskip 1.2 cm

{\large 
Takehiko Asaka$^1$ and Shintaro Eijima$^2$
}

\vskip 0.4cm

$^1${\em
  Department of Physics, Niigata University, Niigata 950-2181, Japan
}

$^2${\em
  Graduate School of Science and Technology, Niigata University, Niigata 950-2181, Japan
}

\vskip 0.4cm

(August 16, 2013)

\vskip 2cm

\vskip .5in

\begin{abstract}
 We consider an extension of the Standard Model by two right-handed 
neutrinos, especially with masses lighter than charged $K$ meson.
This simple model can realize the seesaw mechanism for neutrino 
masses and also the baryogenesis by flavor oscillations of right-handed
neutrinos.
We summarize the constraints on right-handed neutrinos from direct 
searches as well as the big bang nucleosynthesis.
It is then found that the possible range for 
the quasi-degenerate mass of right-handed 
neutrinos is $M_N \geq 163 \MeV$ for normal hierarchy of neutrino
masses, while $M_N = 188 \text{--} 269 \MeV$ and 
$M_N \geq 285 \MeV$ for inverted hierarchy case.
Furthermore, we find in the latter case that the possible value of 
the Majorana phase is restricted for 
$M_N = 188 \text{--} 350 \MeV$, 
which leads to the fact that the rate of neutrinoless double beta
decay is also limited.

\end{abstract}
\end{center}
\end{titlepage}
\renewcommand{\thefootnote}{\#\arabic{footnote}} 
%
\section{Introduction}
\label{sec:introduction}
Various oscillation experiments have revealed non-zero masses of
neutrinos.  The observation shows that there exist two mass scales of
neutrinos, the differences of mass squared $\Delta m^2_{\rm atm}
\simeq 2.43 \times 10^{-3} \eV^2$ and $\Delta m^2_{\rm sol} \simeq 
7.54 \times 10^{-5} \eV^2$ ~\cite{Fogli:2012ua}
, related to the so-called atmospheric and solar neutrinos,
respectively.  In the (canonical) Standard Model neutrinos are exactly
massless, and new physics beyond the Standard Model is indicated.
The crucial questions are then (i) what is the origin of
neutrino masses? and (ii) how do we verify it experimentally?

One of the simplest and most attractive ways to generate neutrino
masses is to introduce right-handed neutrinos $\nu_R$'s into the
Standard Model.  In this case neutrinos can obtain the Dirac masses as
quarks and charged leptons.  Furthermore, since these neutral fermions
are singlets under the gauge group of the Standard Model, the Majorana
masses of right-handed neutrinos are also allowed.  Notice that we
should require at least two right-handed neutrinos in order to explain
$\Delta m^2_{\rm atm}$ and $\Delta m^2_{\rm sol}$.
When the Majorana masses are much heavier than the Dirac
ones, the smallness of neutrino masses can be explained by the seesaw
mechanism~\cite{Seesaw}.  The mass eigenstates are then separated into
three lighter and two heavier ones.
The former ones are responsible to oscillation phenomena
while the latter ones have not bean confirmed by experiments.

In this framework all the neutrinos are Majorana particles,
and the (total) lepton number is violated
which may be tested by the neutrinoless double beta
($0 \nu 2 \beta$) decay. Especially, when only two right-handed 
neutrinos are present, lightest (active) neutrino becomes massless
and the predicted range of the rate for $0 \nu 2 \beta$ decay is
very limited. Interestingly, the future $0 \nu 2 \beta$ experiments 
may probe such a range in the inverted hierarchy of neutrino
masses. (See, for example, a review~\cite{Rodejohann:2012xd})

On the other hand, right-handed neutrinos can also play important
roles to generate the baryon asymmetry of the universe.  The concrete
scenario for this baryogenesis strongly depends on the mass scales of
right-handed neutrinos.  In the canonical leptogenesis
scenario~\cite{Fukugita:1986hr} the lightest right-handed neutrino
should be heavier than ${\cal O}(10^9)$
GeV~\cite{Leptogenesis,Giudice:2003jh} when they have hierarchical
masses.  If $\nu_R$'s are produced non-thermally at the reheating of
the inflation, the required mass can be small as ${\cal O}(10^6)$
GeV~\cite{Leptogenesis_inflation_decay}.  Moreover, the resonant
leptogenesis~\cite{Pilaftsis:2003gt} 
by quasi-degenerate right-handed neutrinos is
possible even for smaller masses of $\nu_R$'s.
Interestingly, the successful scenario of baryogenesis can be realized
even if the masses are smaller than the electroweak
scale~\cite{Akhmedov:1998qx,Asaka:2005pn}.
In these extensions of the Standard Model, therefore, the detailed
examination of right-handed neutrinos is crucial in order to elucidate
the mechanism to generate neutrino masses as well as the cosmic baryon
asymmetry.  For this purpose, the scenario with lighter $\nu_R$'s is
more promising.  

In this paper, we consider the Standard Model with two right-handed
neutrinos which is probably the minimal extension to explain the
neutrino oscillation results and the baryon asymmetry.  Especially, we
assume masses of these neutral leptons are smaller than $K^\pm$
meson.  Such light $\nu_R$'s are good targets of the search
experiments by using $K^\pm$ and $\pi^\pm$ decays (see
Refs.~\cite{Kusenko:2004qc,Gorbunov:2007ak,Atre:2009rg,Asaka:2012hc,Asaka:2012bb}).
It should be noted that, even when $\nu_R$'s are lighter than
$m_{K^\pm}$, $\Delta m^2_{\rm atm}$ and $\Delta m^2_{\rm sol}$ from
the oscillation experiments can be explained via the seesaw mechanism
by requiring that the Yukawa coupling constants of neutrinos are
sufficiently small, and also the enough baryon asymmetry can be
generated via the mechanism~\cite{Akhmedov:1998qx,Asaka:2005pn} by
requiring that the $\nu_R$ 's are sufficiently degenerate.

Under the above situation, we shall summarize the constraints on such
light $\nu_R$'s from the direct search experiments as well as the
cosmological observations, and then identify the allowed region in the
parameter space of the model.  In particular, the possible values of
$\nu_R$ masses are presented, as pointed out by
Ref.~\cite{Gorbunov:2007ak}.

In addition, we shall discuss the implications of the allowed
parameter space.  Especially, we find that the Majorana phase (which
is the one of the CP violating parameter in the lepton sector) is
restricted from the search and cosmological constraints when the
inverted hierarchy of neutrino masses is considered.  It will be then
discussed that this leads to the important impact on the $0\nu 2
\beta$ experiments in future.

\section{Extension by two right-handed neutrinos}
\label{sec:model}
Let us start with the framework of the present analysis.
We consider the Standard Model extended by two right-handed neutrinos
$\nu_R$'s%
\footnote{
By adding the keV right-handed neutrino,
it can play a role of the cosmic dark matter~\cite{Asaka:2005an}.
The results in the present analysis can be applied to the neutrino
Minimal Standard Model ($\nu$MSM)~\cite{Asaka:2005an,Asaka:2005pn} 
if the dark matter physics requires
no limitation to the parameters of the considering two right-handed 
neutrinos.}
together with 
\begin{eqnarray}
  \label{eq:L}
  {\cal L} 
  =
  i \, \overline{\nu_R} \, \gamma^\mu \, \partial_\mu \, \nu_R
  -
  \Bigl(
  F \, \overline{L} \, \Phi \, \nu_R
  + \frac{M_M}{2} \, \overline{\nu_R^c} \, \nu_R
  + h.c.
  \Bigr)
\,,
\end{eqnarray}
where $\Phi$ and $L_\alpha = (e_L, \nu_L)^T$ are Higgs and lepton
doublets, respectively.  Here and hereafter, the indices
of flavor are implicit unless otherwise mentioned.
The $3\times 2$ Yukawa matrix of neutrinos is denoted by $F$
and the $2 \times 2$ matrix of Majorana masses by $M_M$.
Notice that we choose the basis in which the mass matrix of charged
leptons and $M_M$ are diagonal.

We assume that the Dirac masses of neutrinos $M_D = F \langle \Phi
\rangle$ are much smaller than the Majorana ones $M_M$ for the seesaw
mechanism.  In this case, the mass matrix of light neutrinos
participating the observed flavor oscillation is given by $M_\nu = -
M_D M_M^{-1} M_D^T$.  By using this relation, we can parameterize,
without loss of generality, the Yukawa matrix $F$ as
follows~~\cite{Casas:2001sr,Abada:2006ea}
\begin{eqnarray}
  \label{eq:F}
    F = \frac{i}{\vev{\Phi}} \,
    U \, (M_\nu^{\rm diag})^{1/2} \, \Omega \, (M_M)^{1/2} \,,
\end{eqnarray}
where $M_\nu^{\rm diag} = \mbox{diag}(m_1, m_2,m_3)$ 
with masses of light neutrinos $m_i$, and 
$M_\nu^{\rm diag} = U^\dagger M_\nu U^\ast$.
This parametrization is the same as that in Ref.~\cite{Asaka:2011pb}.
The mixing matrix of light neutrinos, called as the
Pontecorvo-Maki-Nakagawa-Sakata (PMNS) matrix~\cite{PMNS},
is written as
\begin{eqnarray}
  U = 
  \left( 
    \begin{array}{c c c}
      c_{12} c_{13} &
      s_{12} c_{13} &
      s_{13} e^{- i \delta} 
      \\
      - c_{23} s_{12} - s_{23} c_{12} s_{13} e^{i \delta} &
      c_{23} c_{12} - s_{23} s_{12} s_{13} e^{i \delta} &
      s_{23} c_{13} 
      \\
      s_{23} s_{12} - c_{23} c_{12} s_{13} e^{i \delta} &
      - s_{23} c_{12} - c_{23} s_{12} s_{13} e^{i \delta} &
      c_{23} c_{13}
    \end{array}
  \right)  
  \times
  \mbox{diag} 
  ( 1 \,,~ e^{i \eta} \,,~ 1) \,,
\end{eqnarray}
with $s_{ij} = \sin \theta_{ij}$ and $c_{ij} = \cos \theta_{ij}$.  

It is then found that the flavor neutrino $\nu_L$ is written 
in terms of mass eigenstates $\nu$'s and $N$'s as
\begin{equation}
 \nu_L = U \nu + \Theta N^C,
\end{equation}
where each heavy neutrino $N$ almost coincides with $\nu_R$, 
$N \simeq \nu_R$.
The mixing elements of $N$ are given by as $\Theta = M_D / M_M$.

In the case under consideration,
there is one Majorana phase $\eta$ in addition to Dirac phase $\delta$,
and the lightest neutrino becomes exactly massless.
Thus, the possible patterns of neutrino masses are 
(i) the normal hierarchy (NH) with
$m_3 > m_2 > m_1 =0$,
and (ii) the inverted hierarchy (IH) with
$m_2 > m_1 > m_3 = 0$.
We express the Majorana mass matrix as
$M_M = {\rm diag} ( M_N - \Delta M/2 \,,~ M_N + \Delta M/2)$
and the $3 \times 2$ matrix $\Omega$ as
\begin{eqnarray}
  \Omega =
  \left(
    \begin{array}{c c}
      0 & 0 \\
      \cos \omega & - \sin \omega \\
      \xi \sin \omega & \xi \cos \omega
    \end{array}
  \right) \,,
\end{eqnarray}
in the NH case, while
\begin{eqnarray}
  \Omega =
  \left(
    \begin{array}{c c}
      \cos \omega & - \sin \omega \\
      \xi \sin \omega & \xi \cos \omega \\
      0 & 0 
    \end{array}
  \right) \,,
\end{eqnarray}
in the IH case.  Here $\omega$ is an arbitrary complex number
and $\xi = \pm 1$ is the sign parameter.
Here we apply the convention $\text{Im}\omega \geq 0$, 
$\xi = \pm 1$ and $\eta = 0 \text{--} \pi$. 

The observational data of mixing angles are $s_{12}^2 =
0.307^{+0.052}_{-0.048}$, $s_{23}^2 = 0.386^{+0.251}_{-0.055}$
 ($0.392^{+0.271}_{-0.057}$), and
$s_{13}^2 = 0.0241^{+0.0072}_{-0.0072}$ 
($0.0244^{+0.0071}_{-0.0073}$), respectively, 
and masses are $\delta m^2 =
m_2^2 - m_1^2 = (7.54^{+0.64}_{-0.55}) \times 10^{-5} \eV^2$ and
$|\Delta m^2| = |m_3^2 - (m_1^2 + m_2^2)/2| = 
(2.43^{+0.19}_{-0.24} )\times 10^{-3}\eV^2$ 
($(2.42^{+0.19}_{-0.25} )\times 10^{-3}\eV^2$)
for the NH (IH) case
(at the $3 \sigma$ level)~~\cite{Fogli:2012ua}.
Hereafter, we shall adopt the central values of these observables.
It should be stressed that these observational data from the
oscillation experiments can be reproduced being independent on the
parameters of right-handed neutrinos, \ie, $M_N$, $\Delta M$, $\omega$
and $\xi$.  In practice, we shall consider the case when $M_N <
m_{K^\pm}$ and $\Delta M \ll M_N$ (see the discussions below).

It is interesting to note that two right-handed neutrinos with $M_N <
m_{K^\pm}$ introduced above can be responsible to the baryon asymmetry
of the universe by invoking the mechanism proposed in
Ref.~\cite{Asaka:2005pn}.  By originating the CP violations in the
flavor oscillation as well as the production/destruction processes of
right-handed neutrinos, the asymmetries of left-handed leptons are
generated which is partially converted into the baryon asymmetry due to
the rapid sphaleron transition~\cite{Kuzmin:1985mm}.  See the analysis
in Refs.~\cite{Asaka:2005pn,Asaka:2010kk,Canetti:2010aw,Canetti:2012vf,
Canetti:2012zc,Canetti:2012kh}.

We perform the numerical study of the generation of the baryon
asymmetry
\footnote{To estimate BAU, we solve numerically 
 the kinetic Eqs.~(5.2) and (5.3) in Ref.~\cite{Asaka:2011wq}
 by neglecting the momentum dependence in the density matrix of 
$\nu_R$'s, for simplicity. The details are found in 
Ref.~\cite{Asaka:2011wq}}
and identify the parameter region in which the observed value
(the baryon to entropy ratio)
$n_B/s = 8.8 \times 10^{-11}$~\cite{Beringer:1900zz}
can be explained.
Here, we use the central values for the neutrino oscillation 
parameters and vary over all the possible value for the 
unknown parameters.

\begin{figure}[tb!]
  \centerline{
    \includegraphics[scale=1.8]{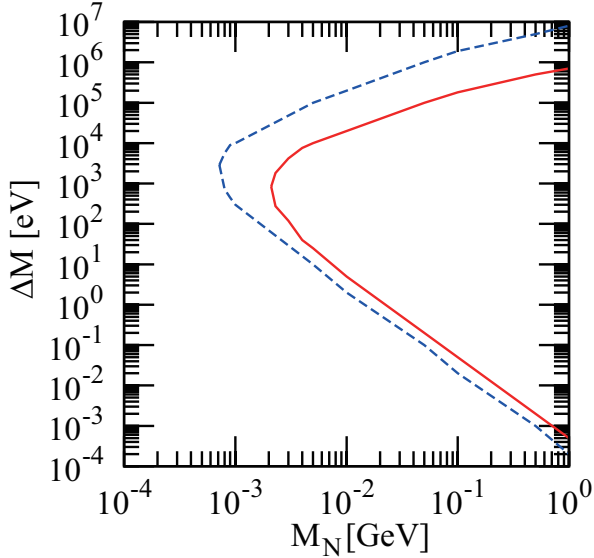}
  }
  \caption{The region of $M_N$ and $\Delta M$ accounting
    for the cosmic baryon asymmetry.
    The observational data can be 
    explained in the regions inside the (red) solid
    and (blue) dashed lines for the NH and IH cases,
    respectively.}
  \label{fig:YB_MN-dM}
\end{figure}

The region for the successful baryogenesis in the 
$M_N$-$\Delta M$ plane is shown in Fig.~\ref{fig:YB_MN-dM}.
It is then found that the enough baryon asymmetry can be generated if
the masses of right-handed neutrinos are
\begin{eqnarray}
  \label{eq:MN}
  M_N \ge
  \left\{
    \begin{array}{l l}
      2.1 \MeV & \mbox{for the NH case}
      \\
      0.7 \MeV & \mbox{for the IH case}
    \end{array}
  \right. \,.
\end{eqnarray}
It is also found that the mass difference of two right-handed neutrinos
should be $\Delta M \ll M_N$.  In these regions, we can explain the
neutrino masses in the oscillation experiments and the baryon asymmetry
at the same time only by using the two right-handed neutrinos.  Such a
mass bound had already been obtained as shown in Fig.~1 of
Ref.~\cite{Canetti:2010aw}.  Our obtained region is wider than theirs,
especially the lower bound on $M_N$ in the NH case is smaller by a
factor of five.  This is mainly because the different values for
$\theta_{i j}$ and $\Delta m_{i j}$ are chosen.

\section{Constraints from direct search and cosmology}
\label{sec:constraint}
The heavy neutrinos ($N \simeq \nu_R$) have the weak interaction due to
the mixing induced by the seesaw mechanism, which strength is suppressed
by the mixing elements $\Theta$ compared with ordinary neutrinos.
Thereby, they can be produced by meson decays 
($\eg$ $\pi \to N e, K \to N \mu$), 
and can decay to Standard Model particles 
($\eg$ $N \rightarrow \nu \nu \bar{\nu}$, $N \rightarrow \nu e^+ e^-$, $N
\rightarrow e \pi$). Using such processes various
experiments have been conducted to search heavy neutrinos directly.
Since such neutrinos have not been discovered, the upper bounds of the
mixing elements have been imposed.  In the discussing mass range, the
beam dump experiment, PS191
experiment~\cite{Bernardi:1985ny,Bernardi:1987ek}, have placed the
strongest bounds on $\Theta$. In this experiment
heavy neutrinos are produced by $\pi$ and/or $K$ decays and charged
particles from $N$ decays in the far detector are searched as signal
events.  These processes are induced as the $\Theta^4$ effect, therefore
the experiment sets the upper bounds of the mixing elements as the
following form, $|\Theta_{\alpha \, I}|^2 \left( a|\Theta_{e \, I}|^2 +
b|\Theta_{\mu \, I}|^2 + c|\Theta_{\tau \, I}|^2 \right)$, where
$\alpha$ and $I$ are the flavor indices of left-handed neutrinos and
heavy neutrinos, respectively. $a$, $b$ and $c$ are coefficients
depending on the mass and decay channel.  In
Refs.~\cite{Bernardi:1985ny,Bernardi:1987ek}, such bounds had been
derived by assuming that heavy neutrino is a Dirac particle and that it
has only the charged current interaction.  As pointed out in
Ref.~\cite{Kusenko:2004qc} (see also Ref.~\cite{Ruchayskiy:2011aa}),
however, the bounds in the considering scenario must be evaluated with
two Majorana (heavy) neutrinos including the neutral current
interaction.

On the other hand, heavy neutrinos are also restricted from 
cosmological observation.
To keep the success of the Big Bang Nucleosynthesis (BBN),  
the lifetime of $N$'s, $\tau_N$, is required 
to be shorter than $0.1$ sec for $M_N > m_\pi$, 
and $\tau_N/\mbox{sec} < t_1 \, (M_N/ \mbox{MeV})^\beta + t_2$
with $t_1 = 128.7$, $t_2 = 0.04179$, and $\beta = - 1.828$
for $M_N \le m_\pi$~\cite{Dolgov:2000jw}. 
Note that the lifetime bound for $M_N \le m_\pi$ is 
discussed recently in Ref.~\cite{Ruchayskiy:2012si}.
Here we apply the bounds in Ref.~\cite{Dolgov:2000jw}
to make the conservative analysis. 

From these constraints we evaluate the possible region of the heavy
neutrinos as performed in previous
work~\cite{Gorbunov:2007ak,Asaka:2010kk,Ruchayskiy:2011aa}.
\begin{figure}[tb!]
  \centerline{
  \includegraphics[scale=1.8]{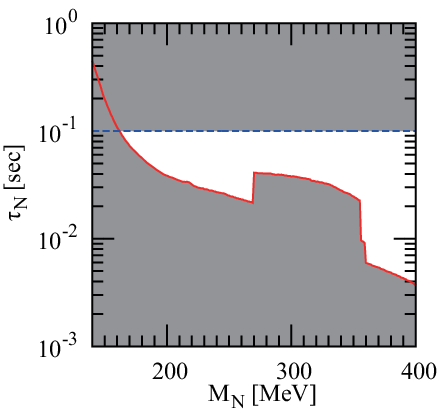}%
  \includegraphics[scale=1.8]{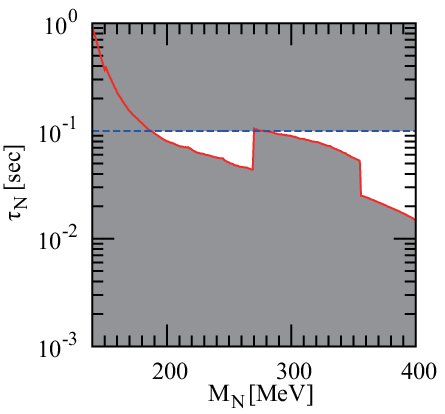}%
   }%
   \caption{The allowed region of $M_N$ and $\tau_N$ in the NH 
            (left panel) and IH (right panel) case, respectively.
            The (red) solid line shows the lower bound on $\tau_N$
            derived from the search experiments, while the (blue) 
            dashed line shows the upper bound on $\tau_N$ by the BBN.}
  \label{fig:LT_K}
\end{figure}
The direct search experiments put the upper bounds on $|\Theta|$ 
which result in the lower bound on $\tau_N$.
On the other hand, the BBN puts the upper bound on $\tau_N$.
Thus, we can identify the possible region of $M_N$ and $\tau_N$.
Fig.~\ref{fig:LT_K} shows the results of this work.
It is found that the following masses of $N$'s are allowed.
\begin{eqnarray}
  \label{eq:MN_ExpBBN}
   \begin{array}{ll}
  M_N \geq 163 \MeV & \mbox{for the NH case}, \\
  M_N = 188 \text{--} 269 \MeV \ \mbox{and} \ M_N \geq \ 285 \MeV 
   &\mbox{for the IH case}.
  \end{array}
\end{eqnarray}

It should be noted that a specific element of the mixing matrix $\Theta$
can be very suppressed compared with other elements by choosing the
parameters carefully~\cite{Asaka:2011pb,Ruchayskiy:2011aa}.  The
possible component of this suppression is $|\Theta_e|^2$ in the NH case,
while that is $|\Theta_\mu|^2$ or $|\Theta_\tau|^2$ in the IH case.  For
the current data from oscillation experiments~\cite{Fogli:2012ua}, the
suppression condition in the NH case cannot be satisfied exactly.
However, $|\Theta_e|^2$ can be smaller by a few orders of magnitude
compared with $|\Theta_\mu|^2$ and $|\Theta_\tau|^2$ near the
suppression point.  In the IH case the suppression condition can be
satisfied even for the current data.  The suppression is, however,
forbidden to escape the constraints from search experiments and the BBN
for $M_N \le 350$ MeV.  On the other hand, the experimental bounds on
$|\Theta_e|^2$ are stronger than that on $|\Theta_{\mu}|^2$ in almost
all interesting mass range.  Due to the feature of mixing elements
$|\Theta|^2$ and the present experimental bounds, the allowed range of
$M_N$ in the NH case is wider than that in the IH case as shown in
Fig.~\ref{fig:LT_K}.

Before closing this section, it should be mentioned that we have
numerically confirmed the enough baryon asymmetry can be generated in
all the allowed regions shown in Fig.~\ref{fig:LT_K}.  As a result, we
have found the parameter space of heavy neutrinos which can explain the
neutrino masses as well as the observed values of BAU without
conflicting with the experimental and cosmological constraints.

\section{Implication to neutrinoless double beta decay}
\label{sec:0nu2beta}
\begin{figure}[tb!]
  \centerline{
  \includegraphics[scale=1.8]{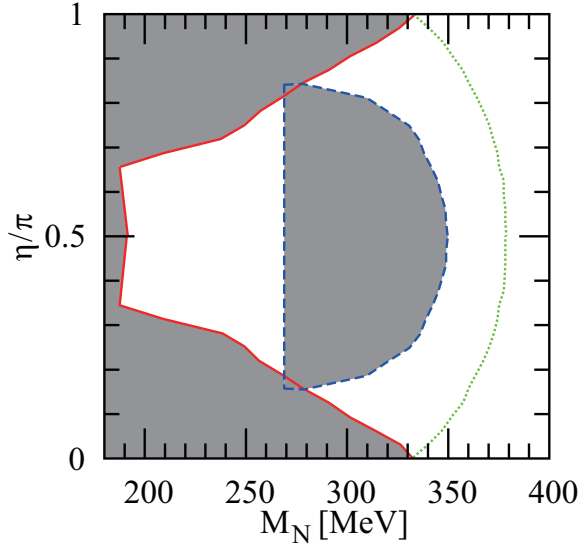}%
   }%
   \caption{The allowed region of $M_N$ and $\eta$ in the IH case.
            The shaded region is excluded by the bounds from
            the search experiments and the BBN.
            The left side of the (red) solid line is excluded by
            using the search mode 
            $K \rightarrow eN \rightarrow e(e\pi)$,
            while the region inside the (blue) dashed line is excluded
            by $K \rightarrow \mu N \rightarrow \mu (\mu \pi)$.
            In the left side of the (green) dotted line only 
            the sign parameter $\xi = +1$ is allowed,
            while both possibility $\xi = \pm1$ is allowed in the 
            opposite side.}
  \label{fig:Majorana}
\end{figure}

In the allowed parameter space, we find a distinctive feature
in the IH case, namely the possible value of the Majorana phase $\eta$
is restricted by the present search experiments and the BBN, as shown in
Fig.~\ref{fig:Majorana}.  In the figure, the shaded region is excluded,
and the (red) solid and (blue) dashed lines are derived by the
experimental bound from the search mode $K \rightarrow eN \rightarrow
e(e\pi)$ and that from the mode $K \rightarrow \mu N \rightarrow \mu
(\mu \pi)$ in PS191, respectively.  The former one puts essentially the
upper bound on $|\Theta_e|^2$ while the latter puts the bound on
$|\Theta_\mu|^2$.  In the left side of the (green) dotted line the sign
parameter $\xi$ is allowed to be only $+1$, while both signs 
$\xi = \pm1$ are allowed in the right side.
It can be seen that, when the mass is $M_N = 188 \text{--} 269$ MeV,
the Majorana phase close to $\pi/2$ is possible and
$\eta \simeq 0$ and $\pi$ are forbidden.
When the mass becomes heavy as $M_N = 285 \text{--} 350$ MeV,
the possible range of $\eta$ is changed, and $\eta \simeq 0$ and $\pi$
as well as $\eta \simeq \pi/2$ are disfavored.
If $N$'s are heavy enough as $M_N > 350$ MeV,
the full range of $\eta$ is consistent with the constraints.  

To understand these outcomes, 
we present the formulae of the mixing elements 
for $X_\omega \equiv e^{\text{Im}\omega} \gg 1$~\cite{Asaka:2011pb}
\begin{eqnarray}
  \nonumber
 |\Theta_e|^2 \eqn\simeq 1.2 \times 10^{-8} \left( \frac{\text{MeV}}{M_N} \right)
                       [1 - 0.93 \, \xi \sin{\eta}] X_{\omega}^2 ,\\ 
 \label{eq:ThetaIH}  
 |\Theta_{\mu}|^2 \eqn\simeq 7.6 \times 10^{-9} \left( \frac{\text{MeV}}{M_N} \right)
                         [1 + 0.90 \, \xi \sin{\eta} 
                          - 0.25 \, \xi \cos{\eta} \sin{\delta}
                          + 0.09 \, \xi \sin{\eta} \cos{\delta}] X_{\omega}^2 ,
                          \hspace{10mm}\\ \nonumber
 |\Theta_{\tau}|^2 \eqn\simeq 5.0 \times 10^{-9} \left( \frac{\text{MeV}}{M_N} \right)
                          [1 + 0.86 \, \xi \sin{\eta}
                           + 0.38 \, \xi \cos{\eta} \sin{\delta}
                           - 0.14 \, \xi \sin{\eta} \cos{\delta}] X_{\omega}^2. 
\end{eqnarray}
Notice that the typical value of $X_\omega$ is ${\cal O}(10)$ in the
allowed region of Fig.~\ref{fig:LT_K}.
We can see from Eq.~(\ref{eq:ThetaIH}) that the mixing elements depend 
crucially on the combination $\xi \sin \eta$.
For $M_N \le 332$ MeV, the constraint on $|\Theta_e|^2$ favors 
$\xi \sin \eta \sim 1$, and hence $\xi = -1$ and also $\eta \sim 0$, $\pi$
are excluded.
In addition, when $M_N = 269 \text{--} 350$ MeV, the constraint on 
$|\Theta_\mu|^2$ favors $\xi \sin \eta \sim -1$, which excludes
the possibility $\eta \sim \pi/2$.

We have found that the possible range of $\eta$ is restricted in the IH
case when $M_N \le 350$ MeV.
We should comment that the similar analysis can be done for the NH case.
However, the present observational data of neutrino oscillation and 
search experiments cannot constrain the value of the Majorana phase.
If the future experiments will improve the data we may also have the 
limitation on $\eta$ even in the NH case.

Next, we turn to discuss the impact on the $0 \nu 2 \beta$ decay
from the result in the IH case.
The decay rate of $0\nu2\beta$ decay is characterized by 
the effective neutrino mass $m_{\rm eff}$
(see, \eg, Refs.~\cite{0nu2beta, Rodejohann:2012xd}).
In the model under consideration, it is given by~\cite{Asaka:2011pb}
\begin{eqnarray}
  \label{eq:meff_nuMSM}
  m_{\rm eff} =
  m_{\rm eff}^\nu
  + 
  \sum_{I} M_I \, \Theta_{e I}^2 \,
  f_\beta(M_I)   
  \,,
\end{eqnarray}
where $m_{\rm eff}^\nu \equiv \sum_{i} m_i U_{e i}^2$ and 
$M_I$ denotes the mass eigenvalue of the $I$th heavy neutrino. 
$f_\beta(M_I)$ is a function which represent the suppression of 
the nuclear matrix element to the contribution of heavy neutrinos.
The function is unity for $M_I \ll 1 \text{GeV}$, 
and decrease as $1 / M_I^2$ for $M_I \gg 1 \text{GeV}$.
The details of $f_\beta$ are described 
in Ref.~\cite{Asaka:2011pb}
(see also Refs.~\cite{Hirsch:1995rf,Blennow:2010th}).

To evaluate $m_{\rm eff}$ we can neglect 
the mass difference of heavy neutrinos
because the observed value of baryon asymmetry 
requires a very small mass difference.
In this case, we can approximate $m_{\rm eff}$ 
in Eq.~(\ref{eq:meff_nuMSM}) as~\cite{Asaka:2011pb}
\begin{eqnarray}
 \label{eq:meff2}
  |m_{\rm eff}| = \left[ 1 - f_\beta(M_N) \right] |m_{\rm eff}^\nu|,
\end{eqnarray}
where
\begin{equation}
\label{eq:meff_nu}
|m_{\rm eff}^\nu| = \cos^2\theta_{13} (m_1^2 \cos^4 \theta_{12} 
+ m_2^2 \sin^4 \theta_{12} 
+ 2 m_1 m_2 \cos^2 \theta_{12} \sin^2 \theta_{12} \cos 2 \eta)
^\frac{1}{2}.
\end{equation}
As shown in this equation,
$|m_{\rm eff}^\nu|$ have a significant dependence on $\eta$.
By varying $\eta$ from $0$ to $\pi$
we find $|m_{\rm eff}^\nu|=(1.82 \text{--} 4.79) \times 10^{-2}$ eV.
This range is indeed for the conventional case when only light 
neutrinos contribute to the $0 \nu 2 \beta$ decay.

\begin{figure}[tb!]
  \centerline{
  \includegraphics[scale=1.8]{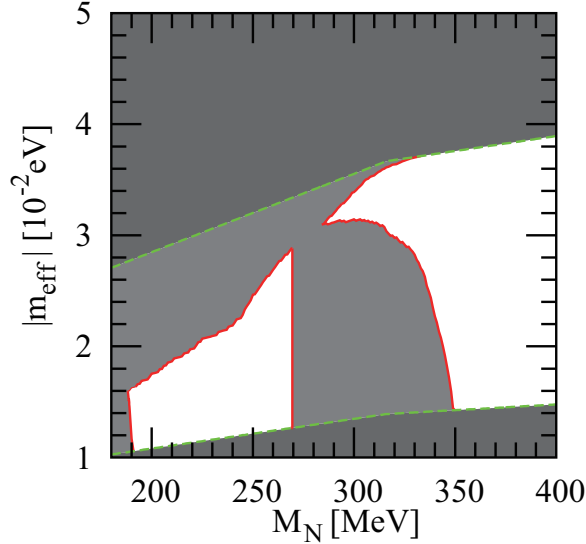}%
   }%
   \caption{The allowed region of $M_N$ and $|m_{\rm eff}|$ in the
            IH case.
            The region between the (green) dashed lines is allowed
            when we take the full range of the Majorana phase
            $\eta$.
            The region between the (red) solid lines is allowed
            by the constraints on $\eta$ from the search
            experiment and the BBN.}
  \label{fig:0nu2beta}
\end{figure}

There are two, important impacts on $|m_{\rm eff}|$ from heavy neutrinos
when the masses are smaller than about $500$ MeV.
The first is the destructive contribution from these particles~\cite{Asaka:2011pb},
which can be easily seen from Eq.~(\ref{eq:meff2}).
The second is the impact from Majorana phase restricted 
by experimental and cosmological constraints.
In Fig.~\ref{fig:0nu2beta} it is shown the predicted range of 
$|m_{\rm eff}|$ being consistent with the constraints on heavy
neutrinos.

We can see the correlation between
the allowed region of $\eta$ in Fig.~\ref{fig:Majorana}
and the predicted range of $|m_{\rm eff}|$ in Fig.~\ref{fig:0nu2beta}.
For $M_N = 188 \text{--} 269$ MeV
$\eta \sim \pi/2$ is favored and then the lower value of $|m_{\rm eff}|$
is predicted, while
$\eta \sim \pi/2$ is disfavored for $M_N = 285 \text{--} 350$ MeV 
and then the larger value of $|m_{\rm eff}|$ is predicted.
When $M_N > 350$ MeV, there is no limitation of $\eta$ but
the presence of heavy neutrinos induces the smaller value of 
$|m_{\rm eff}|$ compared with $|m_{\rm eff}^\nu|$.

\section{Conclusions}
\label{sec:conc}
We have considered right-handed neutrinos which are responsible to
neutrino masses and BAU. Especially, we have discussed the case when
they are quasi-degenerate and lighter than charged kaon.  It has been
found that the constraints from BBN and direct search experiments can be
avoided when the degenerate mass is $M_N \ge 163$ MeV for the NH case
while $M_N = 188 \text{--} 269$ MeV and $M_N \ge 285$ MeV for the IH
case.

Interestingly, we have found that the possible value of Majorana phase
is limited from the present constraints from cosmology and search
experiments if the neutrino masses obey the IH and $M_N = 188 \text{--}
350$ MeV.  In such a case, the rate of $0 \nu 2 \beta$ decay is
also limited which may be different from the prediction by the
conventional scenario with three active neutrinos.  Thus, the future
experiments of $0 \nu 2 \beta$ decay may give us an important hint
for the right-handed neutrinos considered in this analysis.

\section*{Acknowledgments}
The work of T.A. was partially supported by JSPS KAKENHI Grant Number
25400249.



\begin{thebibliography}{999}


\bibitem{Fogli:2012ua} 
  G.~L.~Fogli, E.~Lisi, A.~Marrone, D.~Montanino, A.~Palazzo and A.~M.~Rotunno,
  Phys.\ Rev.\ D {\bf 86}, 013012 (2012)
  [arXiv:1205.5254 [hep-ph]].

\bibitem{Seesaw}
P.~Minkowski,
Phys.\ Lett.\ B {\bf 67} (1977) 421;
T.~Yanagida,
in {\em Proc. of the Workshop on the Unified Theory
and the Baryon Number in the Universe}, 
Tsukuba, Japan, Feb.~13-14, 1979, p.~95, 
eds. O.~Sawada and S.~Sugamoto, 
(KEK Report KEK-79-18, 1979, Tsukuba); 
Progr.\ Theor.\ Phys.\ {\bf 64} (1980) 1103 ; 
M.~Gell-Mann, P.~Ramond and R.~Slansky, 
in {\em Supergravity}, 
eds. P.~van~Niewenhuizen and D.~Z.~Freedman
(North Holland, Amsterdam 1980);
P.~Ramond, 
in {\em Talk given at the Sanibel Symposium}, 
Palm Coast, Fla., Feb.~25-Mar.~2, 1979, preprint CALT-68-709
(retroprinted as hep-ph/9809459);
S.~L.~Glashow,
in {\em Proc. of the Carg\'ese  Summer Institute on Quarks and Leptons},
Carg\'ese, July 9-29, 1979, 
eds. M.~L\'evy et. al, , (Plenum, 1980, New York), p707.

\bibitem{Rodejohann:2012xd} 
  W.~Rodejohann,
  J.\ Phys.\ G {\bf 39}, 124008 (2012)
  [arXiv:1206.2560 [hep-ph]].

\bibitem{Fukugita:1986hr}
  M.~Fukugita and T.~Yanagida,
  Phys.\ Lett.\  B {\bf 174} (1986) 45 .

\bibitem{Leptogenesis}
See, for reviews, \eg,
  W.~Buchmuller, R.~D.~Peccei and T.~Yanagida,
  Ann.\ Rev.\ Nucl.\ Part.\ Sci.\  {\bf 55} (2005) 311
  [arXiv:hep-ph/0502169];
  S.~Davidson, E.~Nardi and Y.~Nir,
  Phys.\ Rept.\  {\bf 466}, 105 (2008)
  [arXiv:0802.2962 [hep-ph]].

\bibitem{Giudice:2003jh}
  For example, see
  G.~F.~Giudice, A.~Notari, M.~Raidal, A.~Riotto and A.~Strumia,
  Nucl.\ Phys.\  B {\bf 685} (2004) 89
  [arXiv:hep-ph/0310123].

\bibitem{Leptogenesis_inflation_decay}
  G.~Lazarides and Q.~Shafi,
  Phys.\ Lett.\ B {\bf 258} (1991) 305;
  H.~Murayama, H.~Suzuki, T.~Yanagida and J.~'i.~Yokoyama,
  Phys.\ Rev.\ Lett.\  {\bf 70} (1993) 1912;
  T.~Asaka, K.~Hamaguchi, M.~Kawasaki and T.~Yanagida,
  Phys.\ Lett.\ B {\bf 464} (1999) 12
  [hep-ph/9906366].
  Phys.\ Rev.\ D {\bf 61} (2000) 083512
  [hep-ph/9907559];

\bibitem{Pilaftsis:2003gt} 
  A.~Pilaftsis and T.~E.~J.~Underwood,
  Nucl.\ Phys.\ B {\bf 692}, 303 (2004)
  [hep-ph/0309342].

\bibitem{Akhmedov:1998qx}
  E.~K.~Akhmedov, V.~A.~Rubakov and A.~Y.~Smirnov,
  Phys.\ Rev.\ Lett.\  {\bf 81} (1998) 1359.

\bibitem{Asaka:2005pn}
  T.~Asaka and M.~Shaposhnikov,
  Phys.\ Lett.\  B {\bf 620} (2005) 17.

\bibitem{Kusenko:2004qc}
  A.~Kusenko, S.~Pascoli and D.~Semikoz,
  JHEP {\bf 0511} (2005) 028
  [arXiv:hep-ph/0405198].

\bibitem{Gorbunov:2007ak}
  D.~Gorbunov and M.~Shaposhnikov,
  JHEP {\bf 0710} (2007) 015
  [arXiv:0705.1729 [hep-ph]].

\bibitem{Atre:2009rg}
  A.~Atre, T.~Han, S.~Pascoli and B.~Zhang,
  JHEP {\bf 0905} (2009) 030
  [arXiv:0901.3589 [hep-ph]].

\bibitem{Asaka:2012hc}
  T.~Asaka and A.~Watanabe,
  JHEP {\bf 1207} (2012) 112
  [arXiv:1202.0725 [hep-ph]].

\bibitem{Asaka:2012bb}
  T.~Asaka, S.~Eijima and A.~Watanabe,
  JHEP {\bf 1303} (2013) 125
  [arXiv:1212.1062 [hep-ph]].

\bibitem{Canetti:2010aw}
  L.~Canetti and M.~Shaposhnikov,
  JCAP {\bf 1009} (2010) 001
  [arXiv:1006.0133 [hep-ph]].


\bibitem{Asaka:2005an}
  T.~Asaka, S.~Blanchet and M.~Shaposhnikov,
  Phys.\ Lett.\  B {\bf 631} (2005) 151.
  [arXiv:hep-ph/0503065].

\bibitem{Casas:2001sr}
  J.~A.~Casas and A.~Ibarra,
  Nucl.\ Phys.\  B {\bf 618} (2001) 171
  [arXiv:hep-ph/0103065].

\bibitem{Abada:2006ea} 
  A.~Abada, S.~Davidson, A.~Ibarra,
  F.~X.~Josse-Michaux, M.~Losada and A.~Riotto,
  JHEP {\bf 0609} (2006) 010
  [arXiv:hep-ph/0605281].

\bibitem{Asaka:2011pb} 
  T.~Asaka, S.~Eijima and H.~Ishida,
  JHEP {\bf 1104}, 011 (2011)
  [arXiv:1101.1382 [hep-ph]].

\bibitem{PMNS}
  B.~Pontecorvo, Sov.\ Phys.\ JETP\ {\bf 7} (1958) 172;\\
  Z.~Maki, M.~Nakagawa and S.~Sakata,
  Prog.\ Theor.\ Phys.\  {\bf 28} (1962) 870.

\bibitem{Kuzmin:1985mm}
  V.~A.~Kuzmin, V.~A.~Rubakov and M.~E.~Shaposhnikov,
  Phys.\ Lett.\  B {\bf 155} (1985) 36.

\bibitem{Asaka:2010kk}
  T.~Asaka and H.~Ishida,
  Phys.\ Lett.\  B {\bf 692} (2010) 105
  [arXiv:1004.5491 [hep-ph]].

\bibitem{Canetti:2012vf} 
  L.~Canetti, M.~Drewes and M.~Shaposhnikov,
  Phys.\ Rev.\ Lett.\  {\bf 110}, no. 6, 061801 (2013)
  [arXiv:1204.3902 [hep-ph]].

\bibitem{Canetti:2012zc} 
  L.~Canetti, M.~Drewes and M.~Shaposhnikov,
  New J.\ Phys.\  {\bf 14}, 095012 (2012)
  [arXiv:1204.4186 [hep-ph]].

\bibitem{Canetti:2012kh} 
  L.~Canetti, M.~Drewes, T.~Frossard and M.~Shaposhnikov,
  Phys.\ Rev.\ D {\bf 87}, 093006 (2013)
  [arXiv:1208.4607 [hep-ph]].

\bibitem{Beringer:1900zz} 
  J.~Beringer {\it et al.}  [Particle Data Group Collaboration],
  Phys.\ Rev.\ D {\bf 86}, 010001 (2012).

\bibitem{Asaka:2011wq}
  T.~Asaka, S.~Eijima and H.~Ishida,
  JCAP {\bf 1202} (2012) 021
  [arXiv:1112.5565 [hep-ph]].


\bibitem{Bernardi:1985ny}
  G.~Bernardi {\it et al.},
  Phys.\ Lett.\  B {\bf 166} (1986) 479.

\bibitem{Bernardi:1987ek}
  G.~Bernardi {\it et al.},
  Phys.\ Lett.\  B {\bf 203} (1988) 332.

\bibitem{Ruchayskiy:2011aa} 
  O.~Ruchayskiy and A.~Ivashko,
  JHEP {\bf 1206}, 100 (2012)
  [arXiv:1112.3319 [hep-ph]].

\bibitem{Dolgov:2000jw}
  A.~D.~Dolgov, S.~H.~Hansen, G.~Raffelt and D.~V.~Semikoz,
  Nucl.\ Phys.\  B {\bf 590}, 562 (2000)
  [arXiv:hep-ph/0008138].

\bibitem{Ruchayskiy:2012si} 
  O.~Ruchayskiy and A.~Ivashko,
  JCAP {\bf 1210}, 014 (2012)
  [arXiv:1202.2841 [hep-ph]].


\bibitem{Hirsch:1995rf}
  M.~Hirsch and H.~V.~Klapdor-Kleingrothaus,
{\it Prepared for International Workshop on Neutrinoless Double Beta Decay and Related Topics, Trento, Italy, 24 Apr - 5 May 1995}  


\bibitem{Blennow:2010th}
  M.~Blennow, E.~Fernandez-Martinez, J.~Lopez-Pavon and J.~Menendez,
  JHEP {\bf 1007} (2010) 096
  [arXiv:1005.3240 [hep-ph]].

\bibitem{0nu2beta}
  F.~Feruglio, A.~Strumia and F.~Vissani,
  Nucl.\ Phys.\  B {\bf 637}, 345 (2002)
  [Addendum-ibid.\  B {\bf 659}, 359 (2003)]
  [arXiv:hep-ph/0201291]; \\
  M.~Hirsch,
  arXiv:hep-ph/0609146.




\end{thebibliography}
\end{document}